\documentclass[aps,prl,10pt,superscriptaddress,twocolumn]{revtex4-2}

\usepackage{mathtools}
\usepackage{amsmath}
\usepackage{amsthm}
\usepackage{amssymb}
\usepackage{tikz-cd}

\usepackage{graphicx}
\usepackage{dcolumn}
\usepackage{bm}
\usepackage{hyperref}
\usepackage{orcidlink}

\usepackage[capitalise,nameinlink]{cleveref}
\usepackage[authormarkuptext=name,commentmarkup=uwave]{changes}
\definechangesauthor[name={AT}, color=red]{alisson}
\definechangesauthor[name={JRH},color=green!70!black]{jonte}
\definechangesauthor[name={NEW}, color=blue!70]{rev}

\newtheorem{lemma}{Lemma}
\newtheorem{proposition}{Proposition}

\newcommand{\af}{\alpha}
\newcommand{\R}{\mathbb{R}}
\newcommand{\ri}{\rightarrow}
\newcommand{\Exp}[1]{\left\langle \psi \left| #1 \psi\right.\right\rangle}
\newcommand{\tra}[1]{\text{tr}\left(#1\right)}
\newcommand{\ketbra}[1]{\vert #1 \rangle \langle #1 \vert}

\begin{document}

\title{Deterministic Underlying States are incompatible with a Counterfactual account of L\"uders' rule}

\author{Alisson Tezzin\,\orcidlink{0000-0002-5849-0124}}
\email{alisson.tezzin@usp.br}
\thanks{Funded by National Council for Scientific and Technological Development (CNPq)}
\affiliation{Department of Mathematical Physics, Institute of Physics, University of São Paulo, R. do Matão 1371, São Paulo 05508-090, SP, Brazil}

\author{B\'arbara Amaral\,\orcidlink{0000-0003-1187-3643}}
\thanks{Funded by São Paulo Research Foundation (FAPESP) and Serrapilheira Institute}
\affiliation{Department of Mathematical Physics, Institute of Physics, University of São Paulo, R. do Matão 1371, São Paulo 05508-090, SP, Brazil}

\author{Jonte R. Hance\,\orcidlink{0000-0001-8587-7618}}
\email{jonte.hance@newcastle.ac.uk}
\thanks{Funded by a Royal Society Research Grant (RG/R1/251590), and an EPSRC Quantum Technologies Career Acceleration Fellowship (UKRI1217).}
\affiliation{School of Computing, Newcastle University, 1 Science Square, Newcastle upon Tyne, NE4 5TG, UK}
\affiliation{Quantum Engineering Technology Laboratories, Department of Electrical and Electronic Engineering, University of Bristol, Woodland Road, Bristol, BS8 1US, UK}

\date{\today}

\begin{abstract}
In this work, we show that a counterfactual account of L"uders’ rule —-- which we argue is naturally implied by the mathematical structure of the rule itself —-- rules out underlying-state models of quantum mechanics (a type of hidden-variable model, typically used in the contextuality and nonlocality literature, where quantum states are treated as probability measures over ``better-defined states''). This incompatibility arises because the counterfactual update requires ontological models to update their states according to conditional probability, which in turn establishes an equivalence between compatibility and the existence of such models.
\end{abstract}

\maketitle

\textit{Introduction---} One of the most debated fundamental problems in modern physics is whether the quantum formalism's description of nature is complete, or if it could be supplemented with additional, or ``hidden'', variables \cite{einstein1935complete, bohr1935complete, bell1966problem, kochen1975problem, bohm1952hiddenI, broglie1960wave, belinfante1973survey}. While hidden-variable models have been constructed which (partially or fully) reproduce the predictions of quantum mechanics \cite{bohm1952hiddenI, bohm1852hiddenII, wiener1955differentialTheory, belinfante1973survey}, various impossibility theorems, notably Bell's \cite{bell1964epr, bell1985beables} and Kochen and Specker's \cite{kochen1975problem}, have shown that such attempts inevitably violate certain consistency conditions which we would arguably want from a successful description of reality \cite{neumann1934algebraic, jauch1963hidden, bell1964epr, kochen1975problem,malley2004commuteSimultaneously}. These theorems differ in the conditions they impose, but a common feature is their indirect dependence on incompatible observables \cite{kochen1975problem, bell1985beables, doring2005vonNeumann, landsman2017foundations}.

In this paper, we explore the quantum state update rule (L\"uders' rule \cite{luders1950upate}) to establish an equivalence between the pairwise compatibility of all observables in a scenario, and the existence of a specific type of deterministic hidden-variable model for that scenario. The models we consider (called deterministic ``underlying-state models'' hereafter) are those in which underlying states determine the values of quantum observables, while quantum states act as statistical mixtures of (i.e., probability measures over) these ``better defined states'' \cite{bell1966problem}. These are the models typically considered in the contextuality and nonlocality literature \cite{bell1966problem,kochen1975problem,amaral2018graph,spekkens2005contextuality,brunner2014bell}. In Ref.~\cite{tezzin2025phdThesis,Tezzin2025Update}, we argue that the result remains valid when the assumption of determinism is dropped, meaning it applies to all ontological models \cite{spekkens2005contextuality}, as well as all other hidden-variable models satisfying factorisability (i.e., the union of statistical independence and no superluminal interaction \cite{Hance2024CounterfactualRestriction}), such as are typically used to test Bell nonlocality \cite{bell1985beables,brunner2014bell}. Since we consider L\"uders' rule, our analysis does not apply --- at least not straightforwardly --- to hidden-variable reformulations of quantum mechanics that (directly \cite{Hance2022MeasProb}) exclude the collapse postulate, such as de Broglie-Bohm pilot wave theory \cite{norsen2017foundations}.

Fine \cite{fine1982bell,fine1982commutingObservables} and Malley \cite{malley2004commuteSimultaneously,malley2005localRealism} have previously argued that incompatibility is sufficient to rule out underlying states. Our work differ from theirs in the consistency conditions we pose: conditional probability (our main assumption) is a theorem in Malley's work \cite{malley2004commuteSimultaneously}, while their consistency conditions follows from ours.

This paper is structured as follows. To begin with, we defend a counterfactual account of L\"uders' rule: instead of being updated by propositions in the simple past (``a measurement of $\hat{A}$ \textit{was} made, and the value $\af$ \textit{was} obtained''), we argue that quantum states are theoretically updated to ensure the validity of propositions in the first conditional (``\textit{if} a measurement is made of $\hat{A}$, the result \textit{will be} found to be $\af$'') \cite{doring2011what}.
This approach aligns with multiple perspectives on quantum mechanics \cite{nakamura1962neumann, bub1977conditionalization, earman2019conditionalization, mermin2022noProblem,mueller2024idealism}. By this account, we argue L\"uders' rule should translate into underlying-state models as conditional probability, because considerations about interactions with measuring apparatuses are automatically removed in the counterfactual description. Using Bayes' rule and Kolmogorov's extension theorem \cite{tao2011measure}, we then show that a set of quantum observables admits a state-updating underlying-state model (where states update via conditional probability) if and only if that set consists of pairwise compatible observables. A single pair of incompatible observables is sufficient to rule out a scenario being representable by a state-updating hidden-variable model. Building on this result, we analyse ontological and metaphysical views typically associated with the absence of hidden variables in quantum mechanics.

\textit{Quantum state update---} We first argue that L\"uders' rule update states based on counterfactual propositions about subsequent measurements, rather than factual statements about a past observation. Note that this is an account of the theoretical quantum update mechanism taken as an axiom in standard quantum mechanics. It should not be misunderstood as a particular interpretation of any physical process which occurs when quantum systems are measured in actual experiments. 

Let $\hat{A}$ be an observable in a finite-dimensional quantum system $\mathfrak{S}$, and let $\af$ be one of its eigenvalues. Consider the counterfactual proposition ``If a measurement is made of $\hat{A}$, the result will be found to be $\af$'' \cite{doring2011what}, denoted $[\hat{A} =\af]$ hereafter. Let $\hat{\Pi}(\hat{A}=\alpha) \equiv \chi_{\{\af\}}(\hat{A})$ be the corresponding projection (in which $\chi_{\{\af\}}$ is the indicator function of $\{\af\}$ and $ \chi_{\{\af\}}(\hat{A})$ is given by the functional calculus \cite{kadison1983fundamentals}), and let $\hat{\Pi}(\hat{A}=\alpha)(\mathcal{H})$ be the subspace of $\mathcal{H}$ onto which $\hat{\Pi}(\hat{A}=\alpha)$ projects, i.e.,
\begin{equation}
    \hat{\Pi}(\hat{A}=\alpha)(\mathcal{H}) = \{\hat{\Pi}(\hat{A}=\alpha)\psi: \psi \in \mathcal{H}\},
\end{equation}
in which $\mathcal{H}$ denotes the finite-dimensional Hilbert space associated with $\mathfrak{S}$.

Pure states that lie within the subspace $\hat{\Pi}(\hat{A}=\alpha)(\mathcal{H})$ ensure the proposition is true, in that they assign probability $1$ to $[\hat{A}=\af]$. This probability is strictly smaller than $1$ for any other pure state. More broadly, a state $\hat{\rho}$ (i.e., a density operator) assigns probability $1$ to $[\hat{A}=\af]$ if and only if it is a convex combination of pure states lying in the corresponding subspace.

The orthocomplement of $\hat{\Pi}(\hat{A} = \alpha)(\mathcal{H})$, namely the set
\begin{equation}
\begin{split}
    \hat{\Pi}&(\hat{A} = \alpha)(\mathcal{H})^{\perp} \doteq\label{eq:orthocomplement}\\
    &\{\phi \in \mathcal{H}: \forall_{\psi \in \hat{\Pi}(\hat{A} = \alpha)(\mathcal{H})}\langle \phi\vert \psi \rangle = 0\},
\end{split}
\end{equation}
is the subspace corresponding the proposition $[\hat{A} \neq \af]$, which asserts that ``If a measurement is made of $\hat{A}$, the result will be found to be different to $\af$''. This is because a pure state $\psi$ satisfies $P_{\psi}[\hat{A} \neq \af]=1$ if and only if $\psi \in \hat{\Pi}(\hat{A} = \alpha)(\mathcal{S})^{\perp}$, in which, for any state $\hat{\rho}$
\begin{equation}
    P_{\hat{\rho}}[A=\af] \equiv \tra{\hat{\rho}\hat{\Pi}(\hat{A}=\af)}
\end{equation}
and $P_{\hat{\rho}}[A\neq\af] \doteq 1-P_{\hat{\rho}}[A=\af]$.

These two subspaces have trivial intersection (i.e., their intersection is the zero vector), and their direct sum is the entire Hilbert space \cite{kadison1983fundamentals}. This means that, for any vector $\psi\in \mathcal{H}$, there exists a unique pair $\psi_{\af} \in \hat{\Pi}(\hat{A} = \alpha)(\mathcal{H})$, $\psi_{\neg\af} \in \hat{\Pi}(\hat{A} = \alpha)(\mathcal{H})^{\perp}$, such that
\begin{equation}
    \psi = \psi_{\af} + \psi_{\neg\af}.
\end{equation}
Furthermore, $\psi_{\af}$ is the unique element of $\hat{\Pi}(\hat{A} = \alpha)(\mathcal{H})$ that minimises the distance from $\psi$, in that
\begin{equation}
    \Vert \psi - \psi_{\af}\Vert =\label{eq:minimizeDistance} \min \{\Vert \phi - \psi\Vert: \phi \in \hat{\Pi}(\hat{A} = \alpha)(\mathcal{H})\},
\end{equation}
and analogously for $\psi_{\neg\af}$. The vector $\psi_{\af}$ is said to be the orthogonal projection of $\psi$ on $\hat{\Pi}(\hat{A} = \alpha)(\mathcal{H})$. It follows by construction that
\begin{equation}
    \psi_{\af} = \hat{\Pi}(\hat{A} = \alpha) \psi.
\end{equation}

Projecting a pure state $\psi$ onto $\hat{\Pi}(\hat{A} = \alpha)(\mathcal{H})$ is the optimal way of reconstructing $\psi$ to ensure the validity of the proposition $[\hat{A} =\af]$, i.e., to ensure --- at the theoretical level --- that, if a measurement is made of $\hat{A}$, the result \emph{will be found to be $\af$}. This is because, among all pure states (which we can consider as unidimensional subspaces) that assign probability 1 to $[\hat{A} = \alpha]$, the subspace spanned by $\hat{\Pi}(\hat{A} = \alpha)\psi$ is the closest to $\psi$, as the very notion of \textit{orthogonal} projection implies. Hence, $\hat{\Pi}(\hat{A} = \alpha)\psi$ is the state in $\hat{\Pi}(\hat{A} = \alpha)(\mathcal{H})$ that best approximates $\psi$ in a mathematically precise sense. As shown in the Appendix, this reasoning naturally extends to the update of density operators.

We denote by $T_{[\hat{A}=\af]}: \mathcal{S}_{0} \ri \mathcal{S}_{0}$ the mapping given by
\begin{equation}
    T_{[\hat{A}=\af]}(\psi) \doteq \frac{\hat{\Pi}(\hat{A}=\af)\psi}{\Vert\hat{\Pi}(\hat{A}=\af)\psi\Vert }
\end{equation}
for each $\psi \in \mathcal{S}_{0}$, where $\mathcal{S}_{0}$ consists of all pure states of the system.

\textit{Order-dependent predictions and incompatibility---} We will denote the distribution of an observable $\hat{B}$ in the state $T_{[\hat{A} =\af]}(\psi)$ by $P_{\psi}[\hat{B} = \cdot \ |\hat{A} =\af]$. That is, for each $\beta \in\sigma(\hat{B})$ we have
\begin{equation}
    P_{\psi}[\hat{B} =\beta|\hat{A} =\af] \equiv P_{T_{[\hat{A} =\af]}(\psi)}[\hat{B} =\beta].
\end{equation}

Let $\hat{A}$ and $\hat{B}$ be (not necessarily compatible) observables, and let $\psi$ be a pure state. For each $\af \in \sigma(\hat{A})$ and $\beta \in\sigma(\hat{B})$, let us define
\begin{equation}
    P_{\psi}[\hat{A}=\af,\hat{B}=\beta] \doteq\label{eq:sequentialDistribution}P_{\psi}[\hat{A}=\af]P_{\psi}[\hat{B}=\beta | \hat{A} = \af]
\end{equation}
\begin{equation}
    =\label{eq:sequentialDistributionQuantum}\langle \psi |\hat{\Pi}(\hat{A}=\alpha)\hat{\Pi}(\hat{B}=\beta)\hat{\Pi}(\hat{A}=\alpha) | \psi \rangle
\end{equation}
(note that $\{\hat{\Pi}(\hat{A}=\alpha)\hat{\Pi}(\hat{B} = \beta)\hat{\Pi}(\hat{A}=\alpha): (\af,\beta) \in \sigma(\hat{A})\times\sigma(\hat{B})\}$ is a POVM). $P_{\psi}[\hat{A}=\af,\hat{B}=\beta]$ can be thought of as the (theoretically constructed) probability of obtaining values $\af$ and $\beta$ by measuring $\hat{A}$ and $\hat{B}$ in sequence.
Compatibility is equivalent to order-independent predictions of sequential measurements:
\begin{lemma}[Compatibility]\label{lemma:compatibility} Let $\mathfrak{S}$ be a finite-dimensional quantum system. Two observables $\hat{A}$ and $\hat{B}$ of $\mathfrak{S}$ are compatible if and only if, for any pure state $\psi$ and values $\af \in \sigma(\hat{A})$, $\beta \in\sigma(\hat{B})$,
\begin{equation}
\begin{split}
    P_{\psi}[\hat{A}=\af,\hat{B}=\beta]\label{eq:bayesLemma} =P_{\psi}[\hat{B} =\beta,\hat{A}=\af].
\end{split}
\end{equation}
\end{lemma}
The proof can be found in the Appendix. Note that, due to Eq.~(\ref{eq:sequentialDistribution}), Eq.~(\ref{eq:bayesLemma}) is formally equivalent to Bayes' rule.

\textit{State-updating Deterministic underlying-state models---} Deterministic underlying-state models, as often used in the contextuality and nonlocality literature \cite{amaral2018graph,budroni2022contextualityReview,spekkens2005contextuality}, are defined as follows:

Associated with a finite-dimensional system $\mathfrak{S}$ is a measurable space $\boldsymbol{\Lambda} \equiv (\Lambda,\mathcal{A})$ \cite{tao2011measure}; the elements of the underlying set $\Lambda$ are called ``hidden variables'', ``hidden states'', or ``underlying states''. Each underlying state determines the value of all physical quantities of the system, thus observables must be represented by real-valued (measurable) functions on $\boldsymbol{\Lambda}$. The function $f_{\hat{A}}$ representing an observable $\hat{A}$ assigns, to each underlying state $\lambda$, the value $f_{\hat{A}}(\lambda)$ that $\hat{A}$ assumes when the system is in the underlying state $\lambda$ \cite{kochen1975problem}. To use Bell's terminology, underlying states are the ``better defined states'' over which states of $\mathfrak{S}$ are averages \cite{bell1966problem}. Each state $\psi$ of $\mathfrak{S}$ therefore defines a probability measure $\mu_{\psi}$ on $\boldsymbol{\Lambda}$. Given any measurable set $\Omega \subset \Lambda$, $\mu_{\psi}(\Omega)$ is the probability that the system is in an underlying state lying in $\Omega$ \cite{kochen1975problem}.

Underlying states assign truth-values to propositions. The proposition $[\hat{A} = \af]$ (``If a measurement is made of $\hat{A}$, the result will be found to be $\af$'') is rendered true by $\lambda$ if $\af =f_{\hat{A}}(\lambda)$ (i.e., if $\af$ is the value possessed by $\hat{A}$ in the underlying state $\lambda$), and false otherwise \cite{isham1998toposI,doring2011what,landsman2017foundations}. This is because, whenever a physical quantity $\hat{A}$ possesses a definite value, a measurement of $\hat{A}$ must reveal, up to operational constraints such as limited accuracy of instruments, this value to the observer: if a quantity holds a value $\af$, the apparatus's reported value must be as close to $\af$ as the accuracy of the instrument permits. This has two important consequences.

First, the (theoretically constructed) probability $P_{\psi}[\hat{A} =\af]$ that a measurement of a quantity $\hat{A}$ returns the value $\af$ is the probability that, at that moment, the system lies in an underlying state that assigns this value to $\hat{A}$. This means
\begin{align}
    P_{\psi}[\hat{A} = \af] &=\label{eq:probabilityModel} \mu_{\psi}(\Omega(\hat{A}=\af)),
\end{align}
where $\Omega(\hat{A}=\af)$ consists of all underlying states in which $[\hat{A}=\af]$ holds, i.e.,
\begin{equation}
    \begin{split}
        \Omega(\hat{A} =\af) &\doteq\label{eq:setProposition} f_{\hat{A}}^{-1}(\{\af\})\\
        &=\{\lambda \in \Lambda: f_{\hat{A}}(\lambda) =\af\}.
    \end{split}
\end{equation}

Second, Bayesian inference (i.e., conditional probability) is the optimal way of reconstructing a state to ensure that a proposition $[\hat{A}=\af]$ holds. If the system is described by a state $\mu_{\psi}$, the proposition $[\hat{A} = \af]$ --- which, under the assumption that underlying states exist, is equivalent to ``the physical quantity $\hat{A}$ has a value, and that value is $\af$'' --- conditions $\mu_{\psi}$ on the set of underlying states for which this claim is true, namely the set $\Omega(\hat{A}=\af)$. Hence, the updated probability measure $\tau_{[\hat{A} = \af]}(\mu_{\psi})$ that reconstructs $\psi$ to ensure that $[\hat{A}=\af]$ holds is given by
\begin{equation}
\begin{split}
    \tau_{[\hat{A} = \af]}(\mu_{\psi})( \ \cdot \ ) &\doteq\label{eq:updateModel} \frac{\mu_{\psi}( \ \cdot \ \cap \Omega(\hat{A}=\af))}{\mu_{\psi}(\Omega(\hat{A}=\af))}
    \\
    &\equiv \mu_{\psi}( \cdot \ |\Omega(\hat{A}=\af)).
\end{split}
\end{equation}

For this to be consistent with the state update mechanism of $\mathfrak{S}$, the probability measure representing $T_{[\hat{A} = \af]}(\psi)$ must be the measure $\mu_{\psi}$ conditioned on $f_{\hat{A}}^{-1}(\{\af\})$, i.e.,
\begin{equation}
    \mu_{T_{[\hat{A} = \af]}(\psi)} =\label{eq:preLudersBayes} \tau_{[\hat{A} = \af]}(\mu_{\psi}).
\end{equation}

To summarise, let $\mathfrak{S}$ be a finite-dimensional quantum system, and let $\mathcal{O}_{S}$ be a non-empty subset of $\mathcal{O}$. A \textbf{state-updating deterministic underlying-state} model for $\mathcal{O}_{S}$ consists of a measurable space $\boldsymbol{\Lambda} \equiv (\Lambda,\mathcal{A})$, a mapping $\Psi$ assigning pure states of $\mathfrak{S}$ to probability measures on $\boldsymbol{\Lambda}$, and a mapping $\Phi$ assigning observables in $\mathcal{O}_{S}$ to measurable functions on $\boldsymbol{\Lambda}$, such that, for each observable $\hat{A} \in \mathcal{O}_{S}$, value $\af \in \sigma(\hat{A})$, and pure state $\psi$, Eqs.~\eqref{eq:probabilityModel} and \eqref{eq:preLudersBayes} are satisfied. This means that the distribution of $\hat{A}$ w.r.t. $\psi$ is that of the random variable $f_{\hat{A}}\equiv \Phi(\hat{A})$ in the probability space defined by $\mu_{\psi} \equiv \Psi(\psi)$ (Eq.~\eqref{eq:probabilityModel}), and that L\"uders' rule translates to $\boldsymbol{\Lambda}$ as conditional probability, as encapsulated by the following commutative diagram
\begin{equation}
        \begin{tikzcd}[column sep=large, arrows={|->}]
        \psi\arrow[r,"\Psi"]\arrow[d,"T_{[\hat{A} = \af]}"] & \mu_{\psi}\arrow[d,"\tau_{[\hat{A} = \af]}"] \\
        T_{[\hat{A} = \af]}(\psi) \arrow[r,"\Psi"] &\mu_{T_{[\hat{A} = \af]}(\psi)}
    \end{tikzcd}
\end{equation}
(as before, $\tau_{[\hat{A} = \af]}$ denotes the mapping that conditions each probability measure $\mu:\mathcal{A} \ri [0,1]$) on $\Omega(\hat{A}=\af)$. When $\mathcal{O}_{S}$ contains all observables of $\mathfrak{S}$, we say that $\mathfrak{M} \equiv (\boldsymbol{\Lambda},\Phi,\Psi)$ is a model for the system $\mathfrak{S}$.

We define an underlying-state model  for a set of observables $\mathcal{O}_{S}$, and not only for the entire system $\mathfrak{S}$, for practical reasons. The subscript $S$ stands for ``scenario'' and is motivated by the concept of a ``measurement scenario'', which appears in operational approaches to quantum foundations \cite{cabello2014graph, brunner2014bell, spekkens2005contextuality}. 

\textit{Incompatibility obstructs state-updating underlying-state models---} We now show that incompatibility obstructs the existence of state-updating underlying-state models. If a deterministic state-updating underlying-state model exists for a set of observables $\mathcal{O}_{S}$, then for any pair of observables $A,B \in \mathcal{O}_{S}$, any pure state $\psi$ and any values $\af \in\sigma(A)$, $\beta \in \sigma(B)$,
\begin{equation}
\begin{split}
    P_{\psi}&[\hat{A} = \af,\hat{B} =\beta] \doteq P_{\psi}[\hat{A} = \af]P_{\psi}[\hat{B} = \beta| \hat{A} = \af]
    \\
    &=  P_{\psi}[\hat{A} = \af]\frac{\mu_{\psi}(\Omega(\hat{A}=\af) \cap \Omega(\hat{B}=\beta))}{P_{\psi}[\hat{A}=\af]}\\
    &=\mu_{\psi}(\Omega(\hat{A}=\af) \cap \Omega(\hat{B}=\beta))\\
    &=P_{\psi}[\hat{B} =\beta]P_{\psi}[\hat{A} = \af|\hat{B} = \beta]\\
    &=P_{\psi}[\hat{B} = \beta,\hat{A} = \af].
\end{split}
\end{equation}
When this is true, according to Lemma~\ref{lemma:compatibility}, $A$ and $B$ are compatible. This proves that a state-updating underlying-state model exists for $\mathcal{O}_{S}$ only if $\mathcal{O}_{S}$ is a set of pairwise compatible observables. On the other hand, the most general version of the Kolmogorov extension theorem (see Theorem 2.4.3 of Ref.~\cite{tao2011measure}) ensures that state-updating underlying-state models can be constructed for any set of pairwise compatible observables (see Appendix for details). This leads to the following result.

\begin{proposition}[Compatibility and underlying states]\label{prop:impossibility} Let $\mathfrak{S}$ be a finite-dimensional quantum system, and let $\mathcal{O}_{S}$ be a non-empty set of observables in $\mathfrak{S}$. The following claims are equivalent.
\begin{itemize}
    \item[(a)] $\mathcal{O}_{S}$ admits a state-updating deterministic underlying-state model.
    \item[(b)] The observables in $\mathcal{O}_{S}$ are pairwise compatible.
\end{itemize}
\end{proposition}

As shown in Ref.~\cite{tezzin2025phdThesis}, a variant of Proposition~\ref{prop:impossibility} also applies to stochastic underlying-state models, whose underlying states assign probabilities --- rather than values --- to observables. It means that, arguably, incompatibility obstructs stochastic underlying-state models as well (or at least those for which state update corresponds to conditional probability \'a la Bayes' rule). We explore the consequences of this result in Ref.~\cite{Tezzin2025Update}. Finally, recall that, regarding contextuality and nonlocality, there is no loss of generality in working with deterministic models \cite{fine1982bell,amaral2018graph,brunner2014bell}.

\textit{Discussion---} Under the counterfactual account of L\"uders' rule, Proposition~\ref{prop:impossibility} provides a straightforward method for ruling out (deterministic) underlying-state models of quantum mechanics. The proof, similar to those in Refs.~\cite{malley2004commuteSimultaneously,malley2005localRealism}, is independent of specific states and observables, as well as space-time considerations; it establishes a direct connection between compatibility and underlying states, demonstrating that incompatibility is sufficient to obstruct such models.

State-updating underlying-state models are inherently Kochen-Specker noncontextual \cite{abramsky2011sheaf,amaral2018graph} and Bell-local \cite{bell1964epr,brunner2014bell}. This means that, if $\vec{A} \equiv (\hat{A}_{1},\dots,\hat{A}_{m})$ are pairwise compatible, then, for any $\vec{\af} \equiv (\af_{1},\dots,\af_{m}) \in \prod_{i=1}^{m}\sigma(\hat{A}_{i})$ and any state $\psi$,
\begin{equation}
    P_{\psi}[\vec{A} = \vec{\af}] =\label{eq:jointDistributionModelM} \mu_{\psi}(\cap_{i=1}^{m}\Omega(\hat{A}_{i}=\af_{i})),
\end{equation}
in which,
\begin{equation}
    \begin{split}
        P_{\psi}[\vec{A} = \vec{\af}] &\doteq\label{eq:jointDistributionM}\\
        \prod_{i=1}^{m} P_{\psi}&[\hat{A}_{i}=\af_{i}|\hat{A}_{1}=\af_{1},\dots,\hat{A}_{i-1}=\af_{i-1}]\\
        &=\Exp{\prod_{i=1}^{m}\hat{\Pi}(\hat{A}_{i}=\af_{i})}
    \end{split}
\end{equation}
(clearly, $P_{\psi}[\hat{A}_{i}= \cdot \ |\hat{A}_{1}=\af_{1},\dots,\hat{A}_{i-1}=\af_{i-1}]$ denotes the distribution of $\hat{A}_{i}$ in the state $(T_{[\hat{A}_{i-1}=\af_{i-1}]} \circ \dots \circ T_{[\hat{A}_{1}=\af_{1}]})(\psi)$). Eq.~(\ref{eq:jointDistributionModelM}) follow from straightforward calculations, as shown e.g., in Ref.~\cite{Tezzin2025Update} (in which it is also proved that state-updating models satisfy Fine's consistency condition \cite{fine1982commutingObservables}). In particular, according to the counterfactual account of L\"uders' rule, contextual deterministic underlying-state models are inconsistent with the canonical postulates of quantum mechanics (which include the collapse postulate).

The view that L\"uders' rule assigns states to \textit{post-measurement states} is particularly popular in the contemporary quantum foundations community, largely due to the influence of quantum information theory \cite{nielsen2000quantum} and the prevalence of operational approaches in the field. This view permeates longstanding debates about the assumptions of ``ideal measurements'', attempts to reconstruct quantum mechanics, and discussions related to hidden variables \cite{nielsen2000quantum, Hindlycke2022ConOntologicalModel, spekkens2007toy, spekkens2005contextuality, budroni2022contextualityReview}. As shown in Ref.~\cite{tezzin2025phdThesis}, the alternative proposed here —-- which agrees with multiple perspectives on quantum mechanics \cite{nakamura1962neumann, bub1977conditionalization, earman2019conditionalization, mermin2022noProblem,muller2023six} --- can shed light on important topics in quantum foundations, extending beyond the problem of hidden variables. 

An important consequence of the correspondence between incompatibility, order-dependent predictions, and constraints on hidden-variable models we establish is that it makes it easier to identify classical parallels to our inability to form  underlying-state models. We will show this in detail in a forthcoming paper. Our work therefore contributes to the debate about whether --- and, if so, why --- phenomena associated with such models, such as contextuality, can be considered a signature of non-classicality \cite{budroni2022contextualityReview,li2017contextualityClassical,li2019independentContextualityClassical,zhang2019ContextualityQuantumClassical,schmid2024inequalities}. 

To conclude, we would like to emphasise that Proposition~\ref{prop:impossibility} corroborates the traditional interpretation of the Kochen-Specker theorem  \cite{kochen1975problem, held2022kochenSpecker, pessoa2006conceitosII, isham1998toposI,  landsman2017foundations,  doring2011what,cabello2022how}: quantum mechanics seems inconsistent with mechanistic realism, or na\"ive realism (as Isham and Butterfield \cite{isham1998toposI}, and D\"oring \cite{doring2011what} put it), where all physical quantities of a system are simultaneously possessed by that system. 

\textit{Acknowledgments---}We thank Rafael Wagner for useful discussions.
\bibliographystyle{apsrev4-2}
\bibliography{Bibliography}

\appendix


\section{Appendix}\label{sec:endMatter}

\noindent \textbf{On the Update of Density Operators}. In the main text, we showed that the update of pure states under objective propositions (those that specify the measurement outcome) is, by construction, a theoretical mechanism for reconstructing states to ensure the validity of certain propositions about future (i.e., subsequent) measurements of physical quantities. Here, we extend this observation to density operators, which represent states where the experimentalist has limited knowledge of the system's pure state.

As we know, the state (i.e., density operator) corresponding to the projected vector $\hat{\Pi}(\hat{A} = \alpha)\psi$ is the rank-1 projection 
\begin{equation} \begin{split}
    T_{[\hat{A} = \af]}(\ketbra{\psi}) &\doteq \left|\frac{\hat{\Pi}(\hat{A} = \alpha)\psi}{\Vert \hat{\Pi}(\hat{A} = \alpha)\Vert}\right\rangle \left\langle\frac{\hat{\Pi}(\hat{A} = \alpha)\psi}{\Vert \hat{\Pi}(\hat{A} = \alpha)\Vert}\right|
    \\
    &=\frac{\hat{\Pi}(\hat{A} = \alpha)\ketbra{\psi}\hat{\Pi}(\hat{A} = \alpha)}{P_{\psi}[\hat{A} = \af]}.
\end{split}\end{equation} 
With a slight abuse of notation, let's denote also by $\mathcal{S}_{0}$ the set of rank-1 projections on $\mathcal{H}$.

The straightforward way of extending the mapping $\mathcal{S}_{0} \ni \ketbra{\psi} \mapsto T_{[\hat{A} = \af]}(\ketbra{\psi}) \in \mathcal{S}_{0}$ to the set $\mathcal{S}$ of all states (i.e., density operators) is by defining
\begin{equation}
    T_{[\hat{A} = \af]}(\hat{\rho}) \doteq\label{eq:firstLuders} \frac{\hat{\Pi}(\hat{A} = \alpha)\hat{\rho}\hat{\Pi}(\hat{A} = \alpha)}{P_{\hat{\rho}}[\hat{A}=\af]}
\end{equation}
for each state $\hat{\rho}$. As well know, this is L\"uders' rule \cite{luders1950upate}, an improvement of von Neumann's collapse postulate \cite{neumann2018foundations}. Let's show that our analysis of the state update rule remains valid under this extension.

As one can easily check, $T_{[\hat{A} = \af]}(\hat{\rho})$ ensures that $[\hat{A}=\af]$ is true, i.e., measurement of $\hat{A}$ in the state $T_{[\hat{A} = \af]}(\hat{\rho})$ yields the value $\af$ with probability $1$ (at the theoretical level). Next, let $\hat{\rho} = \sum_{i=1}^{m} p_{i} \ketbra{\psi_{i}}$ be any convex decomposition of the state $\hat{\rho}$ in terms of  pure states, with $p_{i} >0$ for each $i$ and $\sum_{i=1}^{m} p_{i} = 1$ (recall that $\mathcal{S}$ is the convex hull of $\mathcal{S}_{0}$). The state $\hat{\rho}$ is traditionally interpreted as asserting that the system lies in the state $\ketbra{\psi_{i}}$ with probability $p_{i}$ \cite{nielsen2000quantum}.

In the state $\hat{\rho}$, the probability $P_{\hat{\rho}}[\hat{\rho}=\psi_{i},\hat{A} = \af]$ that the system is in the state $\ketbra{\psi_{i}}$ and that, in this pure state, a measurement of $\hat{A}$ yields the value $\af$ is given by
\begin{equation} \begin{split}
    P_{\hat{\rho}}[\hat{\rho}=\psi_{i},\hat{A} = \af] &\doteq p_{i} P_{\psi_{i}}[\hat{A} =\af].
\end{split}\end{equation}
For simplicity, we will say that $P_{\hat{\rho}}[\hat{\rho}=\psi_{i},\hat{A} = \af]$ is the probability of $[\hat{\rho}=\psi_{i}]$ and $[\hat{A}=\af]$ being ``consecutively true'' in the state $\hat{\rho}$. In the same state, $P_{\hat{\rho}}[\hat{A}=\af]$ is the probability that $[\hat{A}=\af]$ is true. The ratio
\begin{equation}
    \frac{P_{\hat{\rho}}[\hat{\rho}=\psi_{i},\hat{A} = \af]}{P_{\hat{\rho}}[\hat{A}=\af]}
\end{equation}
is thus the probability that $[\hat{\rho}=\psi_{i}]$ and $[\hat{A} =\af]$ are consecutively true, provided that $[\hat{\rho} \in \{\ketbra{\psi_{i}},i=1,\dots,m\} ]$ and $[\hat{A}=\af]$ are satisfied. Now note that
\begin{equation} \begin{split}
    T_{[\hat{A} = \af]}&(\hat{\rho}) =  \sum_{i=1}^{m}p_{i}\frac{\hat{\Pi}(\hat{A} = \alpha)\ketbra{\psi_{i}}\hat{\Pi}(\hat{A} = \alpha)}{P_{\hat{\rho}}[\hat{A}=\af]}
    \\
    &= \sum_{i=1}^{m}\frac{p_{i}P_{\psi_{i}}[\hat{A} =\af]}{P_{\hat{\rho}}[\hat{A}=\af]}T_{[\hat{A} = \af]}(\ketbra{\psi_{i}})\\
    &= \sum_{i=1}^{m}\frac{P_{\hat{\rho}}[\hat{\rho}=\psi_{i},\hat{A} = \af]}{P_{\hat{\rho}}[\hat{A}=\af]}T_{[\hat{A} = \af]}(\ketbra{\psi_{i}}).
\end{split}\end{equation}
Hence, $T_{[\hat{A} = \af]}(\hat{\rho})$ asserts that the system is in one of the states $T_{[\hat{A} = \af]}(\ketbra{\psi_{i}})$, $i=1,\dots,m$, which are pure states ensuring that $[\hat{A}=\af]$  holds. The probability that the system is in the pure state $T_{[\hat{A} = \af]}(\ketbra{\psi_{i}})$ corresponds to the likelihood of $[\hat{\rho}=\psi_{i}]$ and $[\hat{A} =\af]$ being consecutively true, provided that $[\hat{\rho} \in \{\ketbra{\psi_{i}},i=1,\dots,m\} ]$ and $[\hat{A}=\af]$ are satisfied. It shows that our analysis of the state update rule remains valid under L\"uders' extension.

\noindent \textbf{Proof of Lemma~\ref{lemma:compatibility}:}
Let $\hat{A}$ and $\hat{B}$ be (not necessarily compatible) observables in a finite-dimensional quantum system, and let $\hat{A}=\sum_{\af \in \sigma(\hat{A})}\af \hat{\Pi}(\hat{A}=\alpha)$ and $\hat{B}=\sum_{\beta \in \sigma(\hat{B})}\beta \hat{\Pi}(\hat{B} = \beta)$ be their spectral decompositions \cite{landsman2017foundations}. For any pure state $\psi$ and values $\af \in \sigma(\hat{A})$, $\beta \in \sigma(\hat{B})$, we have
\begin{equation}
\begin{split}
    P_{\psi}&[\hat{A} = \af,\hat{B} \in \beta] =\\ 
    &\Exp{\hat{\Pi}(\hat{A}=\alpha)\hat{\Pi}(\hat{B} = \beta)\hat{\Pi}(\hat{A}=\alpha)}.
\end{split}
\end{equation}
On the other hand,
\begin{equation}
\begin{split}
    P_{\psi}&[\hat{B} = \beta,\hat{A} \in \af] =\\
    &\Exp{\hat{\Pi}(\hat{B}=\beta)\hat{\Pi}(\hat{A} = \af)\hat{\Pi}(\hat{B}=\beta)}.
\end{split}
\end{equation}
Pure states separate observables in quantum systems, i.e., two self-adjoint operators $C,D$ are equal if and only if $\Exp{C}=\Exp{D}$ for each state pure $\psi$. Therefore, the probability distributions $P_{\psi}[\hat{A}= \cdot \ ,\hat{B}= \cdot \ ]$ and $P_{\psi}[\hat{B}= \cdot \ ,\hat{A}= \cdot \ ]$  are equal (up to a permutation) for every state $\psi$ if and only if, for all $(\af,\beta) \in \sigma(\hat{A}) \times \sigma(\hat{B})$, the operators $\hat{\Pi}(\hat{A}=\alpha)\hat{\Pi}(\hat{B} = \beta)\hat{\Pi}(\hat{A}=\alpha)$ and $\hat{\Pi}(\hat{B} = \beta)\hat{\Pi}(\hat{A}=\alpha)\hat{\Pi}(\hat{B} = \beta)$ are equal. As shown in Ref.~\cite{rehder1980commute}, this is equivalent to saying that $\hat{\Pi}(\hat{A}=\alpha)$ and $\hat{\Pi}(\hat{B} = \beta)$ are compatible. We know that $\hat{A}$ and $\hat{B}$ are compatible if and only if $\hat{\Pi}(\hat{A}=\alpha)$ and $\hat{\Pi}(\hat{B} = \beta)$ are compatible for each $(\af,\beta) \in \sigma(\hat{A}) \times \sigma(\hat{B})$ \cite{nielsen2000quantum}. Hence, $\hat{A}$ and $\hat{B}$ are compatible if and only if, for each state $\psi$ and each $(\af,\beta) \in \sigma(\hat{A}) \times \sigma(\hat{B})$,
\begin{equation}
    P_{\psi}[\hat{A}=\af,\hat{B}=\beta]=P_{\psi}[\hat{B}=\beta,\hat{A} =\af].
\end{equation}
\hfill\ $\square \medskip$

\noindent \textbf{Proof of Proposition~\ref{prop:impossibility}:} 
     All that remains for us to prove is that (a) follows from (b). Suppose thus that $\mathcal{O}_{S}$ is a set of pairwise compatible observables. Let $\Lambda$ be the Cartesian product $\prod_{\hat{A} \in \mathcal{O}_{S}} \sigma(\hat{A})$, and let $\mathcal{A}$ be the product $\sigma$-algebra $\prod_{\hat{A} \in \mathcal{O}_{S}}\mathfrak{P}(\sigma(\hat{A}))$ \cite{tao2011measure}, where $\mathfrak{P}(\sigma(\hat{A}))$ denotes the collection of all subsets of $\sigma(\hat{A})$. Recall that elements of $\Lambda \equiv \prod_{\hat{A} \in \mathcal{O}_{S}} \sigma(\hat{A})$ are tuples $\lambda \equiv (\lambda_{\hat{A}})_{\hat{A} \in \mathcal{O}_{S} }$ satisfying $\lambda_{\hat{A}} \in \sigma(\hat{A})$ for all $\hat{A} \in \mathcal{O}_{S}$. For each $\hat{A} \in \mathcal{O}_{S}$, let $f_{\hat{A}}$ be the coordinate projection function $\Lambda \ri \sigma(\hat{A})$, that is, $f_{\hat{A}}(\lambda) \doteq \lambda_{\hat{A}}$ for all $\lambda \in \Lambda$. It follows from the definition of product $\sigma$-algebra that $f_{\hat{A}}$ is a measurable function \cite{tao2011measure}. For each $\af \in \sigma(\hat{A})$, denote by $\Omega(\hat{A}=\af)$ the set of all $\lambda \in\Lambda$ such that $\lambda_{A}=\af$, i.e.,  
     \begin{equation}
         \begin{split}
             \Omega(\hat{A}=\af) &\equiv f_{\hat{A}}^{-1}(\{\af\})\\
             &=\{\lambda \in \Lambda: \lambda_{\hat{A}} = \af\}.
         \end{split}
     \end{equation}
    
    Thus far, we have constructed a space of deterministic underlying states $\boldsymbol{\Lambda} \equiv (\Lambda,\mathcal{A})$ and measurable functions $f_{\hat{A}}:\Lambda \ri \R$ representing observables. To conclude, we need to define probability measures representing pure states and recover Eqs.~(\eqref{eq:probabilityModel}) and (\ref{eq:preLudersBayes}). So let $\psi$ be a pure state. For any finite subset $\{\hat{A}_{1},\dots,\hat{A}_{m}\}$ of $\mathcal{O}_{S}$, let 
    \begin{equation}
        P_{\psi}[\vec{A}=\cdot \ ] \equiv P_{\psi}[\hat{A}_{1}=\cdot \ ,\dots,\hat{A}_{m}=\cdot \ ]
    \end{equation}
    be the joint distribution of  $\vec{A} \equiv (\hat{A}_{1},\dots,\hat{A}_{m})$  in the state $\psi$ (see Eq. \ref{eq:jointDistributionM}). Let $\pi$ be any permutation of  $\{1,\dots,m\}$. Then for any $\vec{\af} \equiv (\af_{1},\dots,\af_{m}) \in \prod_{i=1}^{m}\sigma(\hat{A}_{i})$,
    \begin{equation}
        \begin{split}
            P_{\psi}&[\vec{A} = \vec{\af}] = P_{\psi}[\hat{A}_{1}=\af_{1},\dots,\hat{A}_{m}=\af_{m}]\\
            &= \Exp{\prod_{i=1}^{m}\hat{\Pi}(\hat{A}_{i}=\af_{i})}\\
            &=\Exp{\prod_{i=1}^{m}\hat{\Pi}(\hat{A}_{\pi(i)}=\af_{\pi(i)})}\\
            &= P_{\psi}[\hat{A}_{\pi(1)}=\af_{\pi(1)},\dots,\hat{A}_{\pi(m)}=\af_{\pi(m)}].
        \end{split}
    \end{equation}    
    Now let $\hat{A}_{n_{1}},\dots,\hat{A}_{n_{M}}$ be a sub-sequence of $\hat{A}_{1},\dots,\hat{A}_{m}$. The joint distribution of $\hat{A}_{n_{1}},\dots,\hat{A}_{n_{M}}$ is the joint distribution of $\hat{A}_{1},\dots,\hat{A}_{m}$ marginalised over all observables in $\{\hat{A}_{1},\dots,\hat{A}_{m}\}$ except $\hat{A}_{n_{1}},\dots,\hat{A}_{n_{M}}$ --- this is known as the non-disturbance condition \cite{amaral2018graph}. Hence, it follows from the Kolmogorov extension theorem \cite{tao2011measure} that there exists a unique probability measure $\mu_{\psi}$ on $\boldsymbol{\Lambda}$ such that, for any observables $\hat{A}_{1},\dots,\hat{A}_{m} \in \mathcal{O}_{S}$ and values $\vec{\af} \equiv (\af_{1},\dots,\af_{m}) \in \prod_{i=1}^{m}\sigma(\hat{A}_{i})$,
\begin{equation} \begin{split}
    P_{\psi}[\vec{A}=\vec{\af}] &=\label{eq:jointIntersection} \mu_{\psi}\left(\bigcap_{i=1}^{m}f_{\hat{A}_{i}}^{-1}(\{\af_{i}\})\right)
    \\
    &=  \mu_{\psi}\left(\bigcap_{i=1}^{m}\Omega(\hat{A}_{i}=\af_{i})\right)
\end{split}\end{equation}
(see Theorem 2.4.3 of Ref.~\cite{tao2011measure} for details). In particular, for any $\hat{A} \in \mathcal{O}_{S}$ and $\af \in  \sigma(\hat{A})$, Eq.~(\ref{eq:probabilityModel}) is satisfied. Finally, let's show that, for any state $\psi$, any observable $\hat{A}$ and any value $\af\in \sigma(\hat{A})$, the state $T_{[\hat{A} =\af]}(\psi)$ is represented by the measure $\mu_{\psi}$ conditioned on $\Omega(\hat{A}=\af)$, as in Eq.~(\ref{eq:updateModel}). To begin with, it follows from Eqs.~(\ref{eq:jointDistributionM}) and (\ref{eq:jointIntersection})  that, for any observables $\hat{A}_{1},\dots,\hat{A}_{m} \in \mathcal{O}_{S}$ and values $\vec{\af} \equiv (\af_{1},\dots,\af_{m}) \in \prod_{i=1}^{m}\sigma(\hat{A}_{i})$,
\begin{equation} 
\begin{split}
    P_{T_{[\hat{A} =\af]}(\psi)}&[\vec{A}=\vec{\af}] =\\ &\;\;\frac{P_{\psi}[\hat{A} =\af,\hat{A}_{1} =\af_{1},\dots,\hat{A}_{m} =\af_{m}]}{P_{\psi}[\hat{A} =\af]}
    \\
    &=  \frac{\mu_{\psi}( \Omega(\hat{A}=\af) \cap\bigcap_{i=1}^{m}\Omega(\hat{A}_{i}=\af_{i}))}{\mu_{\psi}(\Omega(\hat{A}=\af))}
    \\
    &\equiv \mu_{\psi}(\cap_{i=1}^{m}\Omega(\hat{A}_{i}=\af_{i}) \vert \Omega(\hat{A}=\af)),
\end{split}
\end{equation}
where, as usual, $\mu_{\psi}( \ \cdot \  \vert \Omega(\hat{A}=\af))$ denotes the measure $\mu_{\psi}$ conditioned on $\Omega(\hat{A}=\af)$. On the other hand, Eq.~(\ref{eq:jointIntersection}) entails that 
\begin{equation}
    P_{T_{[\hat{A} =\af]}(\psi)}[\vec{A}=\vec{\af}] = \mu_{T_{[\hat{A} =\af]}(\psi)}(\cap_{i=1}^{m}\Omega(\hat{A}_{i}=\af_{i})),
\end{equation}
therefore
\begin{equation}
\begin{split}
    P_{T_{[\hat{A} =\af]}(\psi)}&[\vec{A}=\vec{\af}]= \mu_{T_{[\hat{A} =\af]}(\psi)}(\cap_{i=1}^{m}\Omega(\hat{A}_{i}=\af_{i}))\\
    &= \mu_{\psi}\left(\bigcap_{i=1}^{m}\Omega(\hat{A}_{i}=\af_{i}) \vert \Omega(\hat{A}=\af)\right).
\end{split}
\end{equation}
Hence, it follows from the uniqueness of the measure given by the Kolmogorov extension theorem \cite{tao2011measure} (applied to the state $T_{[\hat{A} =\af]}(\psi)$) that
\begin{equation}
    \mu_{T_{[\hat{A} =\af]}(\psi)}( \ \cdot \ ) = \mu_{\psi}( \ \cdot \ \vert \Omega(\hat{A}=\af)).
\end{equation}
It proves that $\tau_{[\hat{A} =\af]}$ is necessarily given by Eq.~(\ref{eq:updateModel}), completing the proof.
\hfill\ $\square \medskip$

\end{document}